\documentclass[
  journal=pasa,
  manuscript=research-paper, 
  year=202X,
  volume=YY,
]{cup-journal}

\usepackage{amsmath}
\usepackage{amssymb,microtype,siunitx,booktabs}
\usepackage{multirow}

\title{First detections of methanol maser lines from a rare transition family}

\author{B. R. Johnson}
\affiliation{School of Natural Sciences, University of Tasmania, Hobart, 7005, Tasmania, Australia}
\email[B. R. Johnson]{Bradley.Johnson@utas.edu.au}

\author{S. P. Ellingsen}
\affiliation{School of Natural Sciences, University of Tasmania, Hobart, 7005, Tasmania, Australia}
\alsoaffiliation{International Centre for Radio Astronomy Research, The University of Western Australia, Crawley, WA 6009, Australia}

\author{S. L. Breen}
\affiliation{SKAO, Jodrell Bank, Lower Withington, Macclesfield, Cheshire SK11 9FT, UK}

\author{M. A. Voronkov}
\affiliation{CSIRO Space \& Astronomy, PO Box 76, Epping, NSW 1710, Australia}

\author{T. P. McCarthy}
\affiliation{School of Natural Sciences, University of Tasmania, Hobart, 7005, Tasmania, Australia}

\author{L. J. Hyland}
\affiliation{School of Natural Sciences, University of Tasmania, Hobart, 7005, Tasmania, Australia}

\received {dd Mmm YYYY}
\revised  {dd Mmm YYYY}
\accepted {dd Mmm YYYY}
\published{dd Mmm 202X}

\keywords{masers – stars: formation – ISM: molecules – radio lines: ISM} 

\begin{document}

\begin{abstract}
We report the first observations in a rare family of class II methanol maser transitions in both CH$_3$OH and $^{13}$CH$_3$OH toward three southern high-mass star formation regions, along with the first maser detected in the $^{13}$CH$_3$OH line. The $8_2 \rightarrow 9_1  A^{-}$ methanol transition was observed in both CH$_3$OH and $^{13}$CH$_3$OH (at 28.9 GHz and 41.9 GHz respectively) toward three sources; G358.93-0.03, NGC6334I and G345.01+1.79, all of which are star formation regions with recent maser flaring events. We report the first maser detection of the 41.9 GHz line in $^{13}$CH$_3$OH toward G358.93-0.03 and the first confirmed maser detection of the 28.9 GHz line in CH$_3$OH toward NGC6334I. Additionally we report a maser detection of the 28.9 GHz line in CH$_3$OH toward G358.93-0.03, meaning that with our detection of the 41.9 GHz line, this is the first isotopic detection of these lines toward G358.93-0.03. The newly detected maser transitions are associated with the primary millimetre continuum sources (MM1) in both G358.93-0.03 and NGC6334I, within the varying positional uncertainties.
\end{abstract}

\section{INTRODUCTION}
\label{sec:introduction}

There is much that we don't know about the complex processes that occur during high-mass star formation \citep{Zinnecker:2007,Motte:2018,Beuther:2025}. However, key tracers of star formation activity such as astrophysical methanol masers have greatly increased our understanding of star formation regions and their physical constraints \citep{Breen:2012,Urquhart:2013}.

Methanol masers have been empirically categorised into two different classes \citep{Batrla:1987} depending on the pumping mechanism that creates them; class I methanol masers are collisionally pumped while class II methanol masers are radiatively pumped \citep{Cragg:1992,Sobolev:1994}. The collisionally pumped class I methanol masers are primarily associated with outflows and expanding HII regions \citep{Kurtz:2004,Voronkov:2014}. Whereas the radiatively pumped class II methanol masers are closely associated with bursts of activity around young massive stellar objects and therefore play an important role in the study of high-mass star formation \citep{Minier:2001,Ellingsen:2006}. The strongest and most commonly detected of the class II methanol masers is the 6.7 GHz transition, which is known to be exclusively associated with young high-mass stars \citep{Breen:2013}. 

\subsection{G358.93-0.03 2019 Flare Event}
\label{sec:g358 flare event}

G358.93-0.03 was first identified as a region hosting young, high-mass stars with the detection of the 6.7 GHz methanol maser in March 2006 \citep{Caswell:2010}. Since this discovery the 6.7 GHz maser has maintained a quiescent flux density of <10 Jy \citep{Sugiyama:2019}. In January 2019, a flare of activity was reported toward G358.93-0.03 by the Maser Monitoring Organisation (M2O) \citep{Sugiyama:2019}. The flare was characterised by a steady increase in the peak intensity of the known 6.7 GHz line from $\sim$10 Jy in early January to $\sim$100 Jy by the end of the month, along with several other spectral features also increasing in strength \citep{Sugiyama:2019}. Many astronomers around the world subsequently undertook their own observations of this flaring event \citep{Breen:2019,Brogan:2019,MacLeod:2019,Chen:2020a,Chen:2020b,Volvach:2020,Miao:2022}. The 6.7 GHz line continued to show higher than normal variability over the following months and reached an extreme of 1156 Jy in March 2019 \citep{MacLeod:2019}.
These subsequent observations presented a wealth of new detections of rare maser transitions with many being first ever detections \citep{Breen:2019,McCarthy:2023}. These new observations have also been used to characterise G358.93-0.03 as a late-stage star forming core with accretion arms \citep{MacLeod:2019,Chen:2020b}. 

\subsection{Observed Methanol Transitions}
\label{sec:observed transitions}

\begin{table*}[t]
\caption{$8_2 \rightarrow 9_1  A^{-}$ transition information and maser detection history for CH$_3$OH and $^{13}$CH$_3$OH, including the first detections as a maser; in any source and in each source studied here (N/A means the transition has not been detected as a maser in that source).}
\label{transition info}
\resizebox{\textwidth}{!}{
    \begin{tabular}{ccccccc}
    \toprule
    \headrow Molecule & Spectral Line & First maser detection in any source & \multicolumn{3}{c}{First maser detections in listed sources}\\
    & & & G358.93-0.03 & NGC6334I & G345.01+1.79\\ 
    \midrule
    CH$_3$OH & 28.9 GHz & \cite{Wilson:1993} [W3(OH)] & \cite{Miao:2022} & \cite{Wu:2023} (As a maser candidate) & N/A\\ 
    \midrule
    $^{13}$CH$_3$OH & 41.9 GHz & This paper [G358.93-0.03] & This paper & N/A & N/A\\
    \bottomrule
    \end{tabular}
}
\end{table*}

Here we analyse new data, collected during the 2019 flare event toward G358.93-0.03, in the rare family of class II methanol maser transitions ($8_2 \rightarrow 9_1  A^{-}$) in both CH$_3$OH (methanol) and $^{13}$CH$_3$OH (methanol-13C). Maser emission in the $8_2 \rightarrow 9_1  A^{-}$ transition in CH$_3$OH, at 28.9 GHz, toward G358.93-0.03 was reported for the first time in \cite{Miao:2022}. The only previous maser detection of the 28.9 GHz line was made by \cite{Wilson:1993} toward W3(OH) with a peak flux density of 15 Jy. This transition has also been detected as a thermal emission in Orion KL \citep{Wilson:1993} and NGC 7538 IRS1 \citep{Shuvo:2021} and as a maser candidate toward NGC6334I \citep{Wu:2023}. The equivalent $8_2 \rightarrow 9_1  A^{-}$ transition in $^{13}$CH$_3$OH has a rest frequency of 41.9 GHz and has never before been reported as a maser emission \citep{Kuiper:1989,Wu:2023}.

\subsection{Additional Observations (NGC6334I \& G345.01+1.79)}
\label{sec:additional sources}

In addition to the observations toward G358.93-0.03, the 28.9 GHz and 41.9 GHz transitions were also observed toward two other star formation regions; NGC6334I and G345.01+1.79. The maser detection history for all three sources in the $8_2 \rightarrow 9_1  A^{-}$ transitions of both CH$_3$OH and $^{13}$CH$_3$OH are listed in Table~\ref{transition info}.

The first of these two additional regions, NGC6334I is a well studied star formation region toward which many methanol maser transitions have been observed \citep{Cragg:2001}. NGC6334I began a period of flaring in 2015, resulting in significant changes of the 6.7 GHz line and leading to the observation of many more methanol maser transitions \citep{Hunter:2017}. Previously, the well-known ultracompact HII (UCHII) region NGC6334I-MM3 (also known as NGC6334F) dominated the bright maser emission within the NGC6334I region \citep{Hunter:2018}. However, since the beginning of the flare in 2015 the hot core region NGC6334I-MM1 has shown a large increase in maser activity \citep{Hunter:2017, Brogan:2018, Kumar:2025}.

The southern sky source G345.01+1.79 has previously shown a range of methanol maser transitions \citep{Ellingsen:2012}. The 12.2 and 6.7 GHz class II methanol transitions were initially discovered towards this source in \cite{Norris:1988,Norris:1993}, respectively. Since then this source has been included in multiple maser searches and many more transitions have been detected, suggesting star formation activity \citep{Cragg:2001,Ellingsen:2012,Voronkov:2014}.

\subsection{Previous Isotopic Methanol Maser Detections}
\label{sec:previous isotopic detections}

\cite{Chen:2020a} reported the first detection of isotopic $^{13}$CH$_3$OH maser emission (i.e., detecting masers in the same transition in different isotopologues), in which they detected the $2_0 \rightarrow 3_{-1} E$ transition of both CH$_3$OH and $^{13}$CH$_3$OH (at 12.178 and 14.782 GHz respectively) toward G358.93-0.03 during the 2019 flare event. These detections were made using the Tianma 65-m Radio Telescope (TMRT), the Very Large Array (VLA) and the Australia Telescope Compact Array (ATCA), and also included observations of $^{13}$CH$_3$OH in the $5_1 \rightarrow 6_0 A^+$ (14.300 GHz) and $9_2 \rightarrow 10_1 A^+$ (35.171 GHz) transitions, and CH$_3$OH in the $16_5 \rightarrow 17_4 E$ (12.229 GHz) transition \citep{Chen:2020a}.

\section{OBSERVATIONS AND METHODS}
\label{sec:observations methods}

\begin{table*}[t]
\caption{$8_2 \rightarrow 9_1  A^{-}$ transition information and parameters for the observations of the spectral lines in CH$_3$OH and $^{13}$CH$_3$OH, including the rest frequency (with uncertainties in parenthesis as the last digits), the reference for the rest frequency values, the date of observation, the ATCA configuration, the velocity resolution and the synthesised beam size.}
\label{observation info}
\resizebox{\textwidth}{!}{
    \begin{tabular}{ccccccc}
    \toprule
    \headrow Molecule & Rest frequency & Rest frequency reference & Observation Date & ATCA Configuration & $V_{res.}$ & Synth. Beam\\
    \headrow & [GHz] & & & & [km/s] & ($'' \times ''$) \\
    \midrule
    CH$_3$OH & 28.969942(50) & \cite{Muller:2004} & 2019/04/30 & 750C & 0.323 & 17.0 x 2.3\\ 
    \midrule
    $^{13}$CH$_3$OH & 41.904332(25) & \cite{Xu:1997} & 2019/04/26 & 750C & 0.224 & 7.7 x 1.5\\
    \bottomrule
    \end{tabular}
}
\end{table*}

The data in this paper were collected using the Australia Telescope Compact Array (ATCA) in April 2019 during the flare event toward G358.93-0.03 (see Section~\ref{sec:g358 flare event}). The ATCA was in the 750C configuration and using all six antennas for both sets (28.9 and 41.9 GHz) of observations (see Table~\ref{observation info} for the full observational parameters). In this configuration the five movable antennas are in a line along the middle of the east-west track while antenna six is 3 km west of the main track. This configuration has a minimum baseline length of 46 metres and a maximum baseline length of 5020 metres. Since all six antennas were in an east-west line, this configuration relies on the rotation of the Earth over time to get complete $uv$ space coverage. Therefore for shorter observations, like the observations of the 28.9 GHz line here (Table~\ref{full transition results}), the synthesised beam is highly elongated in declination due to the limited Earth rotation (Table~\ref{observation info}). This inhibits the accurate derivation of spatial coordinates of the maser.

The Compact Array Broadband Backend (CABB; \citealt{Wilson:2011}) was in the 64M-32k mode, which enabled zoom bands of 64 MHz with 2048 channels each, resulting in native velocity resolutions of 0.323 and 0.224 km/s for the 28.9 and 41.9 GHz transitions respectively. Throughout both sets of observations, the phase calibrator source 1714-336 was observed periodically and 1934-638 was used as a flux density calibrator. For bandpass calibration, 1921-293 and 1253-055 were used for the 28.9 GHz and 41.9 GHz transitions respectively.

The initial data were collected on 2019 April 26, and included zoom bands targeting many methanol maser transition lines toward the three sources listed in Section~\ref{sec:additional sources} (ATCA project code CX429, see results from other transitions in \citealt{Breen:2019}). This initial set of observations included a search for the $8_2 \rightarrow 9_1  A^{-}$ transition in the less common $^{13}$CH$_3$OH isotopologue of methanol, at a frequency of 41.9 GHz.
Preliminary analysis of these data showed that the $8_2 \rightarrow 9_1  A^{-}$ transition line for $^{13}$CH$_3$OH was detected during this search. 

A follow-up observation was then conducted on 2019 April 30 which detected the equivalent $8_2 \rightarrow 9_1  A^{-}$ transition in CH$_3$OH at 28.9 GHz. The justification for this follow-up observation is that modelling shows that $^{13}$CH$_3$OH masers can have comparable brightnesses to CH$_3$OH masers in some situations (although in reality $^{13}$CH$_3$OH has an abundance 1/30 that of CH$_3$OH, resulting in weaker maser signals; \citealt{Johns:1998}), therefore the discovery of the $8_2 \rightarrow 9_1  A^{-}$ transition in $^{13}$CH$_3$OH suggests there may be an accompanying transition in CH$_3$OH. However, the 28.9 GHz line is well outside the nominal frequency range of the 7mm receiver system of the ATCA ($\sim$30 GHz to $\sim$50 GHz), which is why it was not included in earlier observations. This presented issues in reducing the data such as a degraded bandpass, which hampered calibration and means only a small frequency range around the signal had acceptably low noise.
The $^{13}$CH$_3$OH transition (at 41.9 GHz) is well within this optimal frequency range and is therefore not affected by this degradation.

The data were reduced using MIRIAD \citep{Sault:1995}. A bandpass model was created for each bandpass calibrator which was then passed to the secondary calibrators for phase and flux density calibration, with the final calibration then being applied to each source. Due to the relatively small number of antennas and the short observation time, which means lower sensitivity, the absolute uncertainty in primary flux density calibration would normally be expected to be $\sim$10$\%$ \citep{Breen:2019}, however due to the detections being outside the nominal band, this uncertainty is likely higher for the 28.9 GHz line.

\section{RESULTS}
\label{sec:results}

\subsection{G358.93-0.03}
\label{sec:g358 results}

\begin{table*}[t]
\begin{threeparttable}
\caption{Properties of the $8_2 \rightarrow 9_1  A^{-}$ transitions in CH$_3$OH and $^{13}$CH$_3$OH toward G358.93-0.03, NGC6334I, and G345.01+1.79; including the date of observation, the minimum, maximum and peak velocities, uncertainty in velocity (based on the uncertainty in the respective rest frequencies), the peak and
integrated flux densities, the RMS noise of a signal-free area of the spectra, and the total integration time of the observation. Note the missing values for many of the properties in G345.01+1.79, as no signal was detected toward this source.}
\label{full transition results}
\begin{tabular}{lllllllllll}
\toprule
\headrow Source & Molecule & Spectral Line & $V_{min}$ & $V_{max}$ & $V_{peak}$ & $V_{Uncert.}$ & $S_{peak}$ & $S_{int}$ & $S_{rms}$ & Int. time\\
\headrow & & & [km/s] & [km/s] & [km/s] & [km/s] & [Jy] & [Jy km/s] & [Jy] & [minutes]\\
\midrule
\multirow{1}{*}{G358.93-0.03}
    & CH$_3$OH & 28.9 GHz & -18.5 & -15.0 & -16.3 & 0.5 & 10.5 & 11.4 & 0.4 & 40\\ 
    & $^{13}$CH$_3$OH & 41.9 GHz & -17.9 & -15.5 & -17.3 & 0.2 & 0.4 & 0.3 & 0.004 & 91\\
\midrule
\multirow{1}*{NGC6334I}
    & CH$_3$OH & 28.9 GHz & -11.4 & -9.5 & -10.5 & 0.5 & 6.2\tnote{a} & 4.1\tnote{a} & 0.4\tnote{a} & 29\\ 
    & $^{13}$CH$_3$OH & 41.9 GHz & -9.6 & 4.7 & -6.5 & 0.2 & 0.2 & 0.8 & 0.004 & 46\\
\midrule
\multirow{1}*{G345.01+1.79}
    & CH$_3$OH & 28.9 GHz & --- & --- & --- & 0.5 & <9.8 & --- & 2.0 & 30\\ 
    & $^{13}$CH$_3$OH & 41.9 GHz & --- & --- & --- & 0.2 & <0.04 & --- & 0.007 & 45\\
\bottomrule
\end{tabular}
\begin{tablenotes}[hang]
\item[a] The value of $S_{peak}$ for NGC6334I was determined by scaling the amplitude of the signal, as described in Section~\ref{sec:ngc6334 results}
\end{tablenotes}
\end{threeparttable}
\end{table*}

\begin{figure}[hbt!]
\centering
\includegraphics[width=0.75\linewidth]{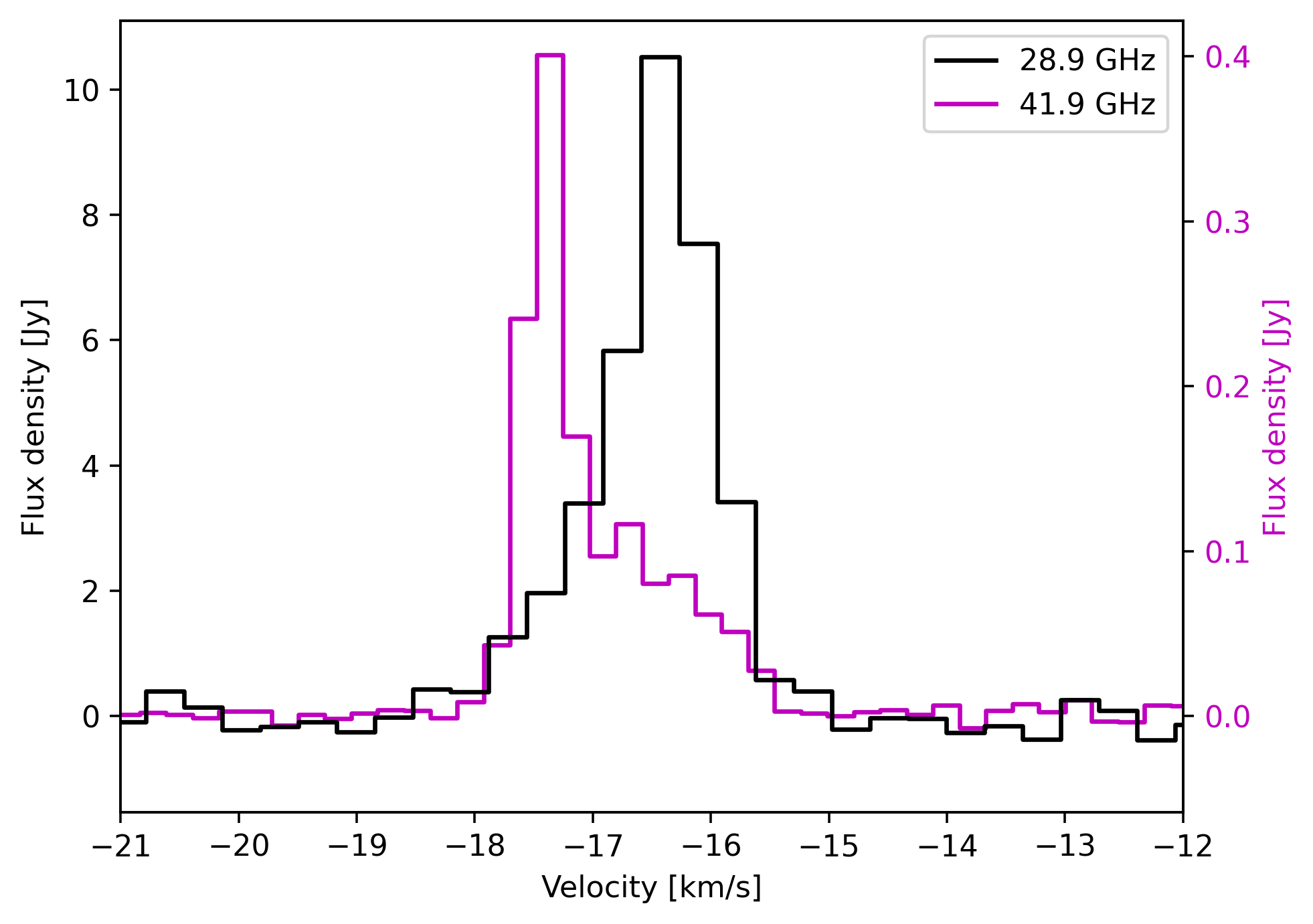}
\caption{G358.93-0.03 Spectra Comparison; The signal in black indicates the 28.9 GHz line detected on 2019 April 30 with a peak of 10.5 Jy, while the magenta line indicates the 41.9 GHz line detected on 2019 April 26 with a peak of 0.4 Jy. Note the difference in the scale of the respective axes; the 28.9 GHz line is $\sim$26 times stronger than the 41.9 GHz line. The similarity in signal-to-noise despite the different scales is due to the longer integration time and the better bandpass in the 41.9 GHz observations}
\label{g358_comparison_figure}
\end{figure}

Toward G358.93-0.03, the 28.9 GHz class II methanol line in CH$_3$OH was detected with a peak flux density of 10.5 Jy at a velocity of -16.3 km/s (Figure~\ref{g358_comparison_figure}). This spectrum aligns with the spectrum of the same line in \cite{Miao:2022} which is broadened toward the lower velocities and shows a primary peak around -16 km/s and multiple smaller peaks between -18 and -16 km/s.

As noted above, we report the first ever detection of maser emission in the 41.9 GHz transition of $^{13}$CH$_3$OH. Our detection of this line (Figure~\ref{g358_comparison_figure}) shows a peak of 0.4 Jy at -17.3 km/s. There is also a broadening of the line toward the higher velocities, which may imply multiple smaller components between -17 and -15 km/s. The peaks of both the 28.9 GHz and 41.9 GHz lines are offset from each other by about 1 km/s and the amplitude of the 28.9 GHz line is a factor of $\sim$26 larger than the 41.9 GHz line.

Due to the rapid changes in amplitude in some class II methanol masers during this flare event, it must be noted that even the short time between the observations of the 28.9 GHz and 41.9 GHz lines (4 days) could have affected the strength of these lines, and strong conclusions should not be drawn from the comparison of their relative flux densities. The full properties of the masers detected toward G358.93-0.03 are presented in Table~\ref{full transition results}.

\subsection{NGC6334I}
\label{sec:ngc6334 results}

\begin{figure}[hbt!]
\centering
\includegraphics[width=0.75\linewidth]{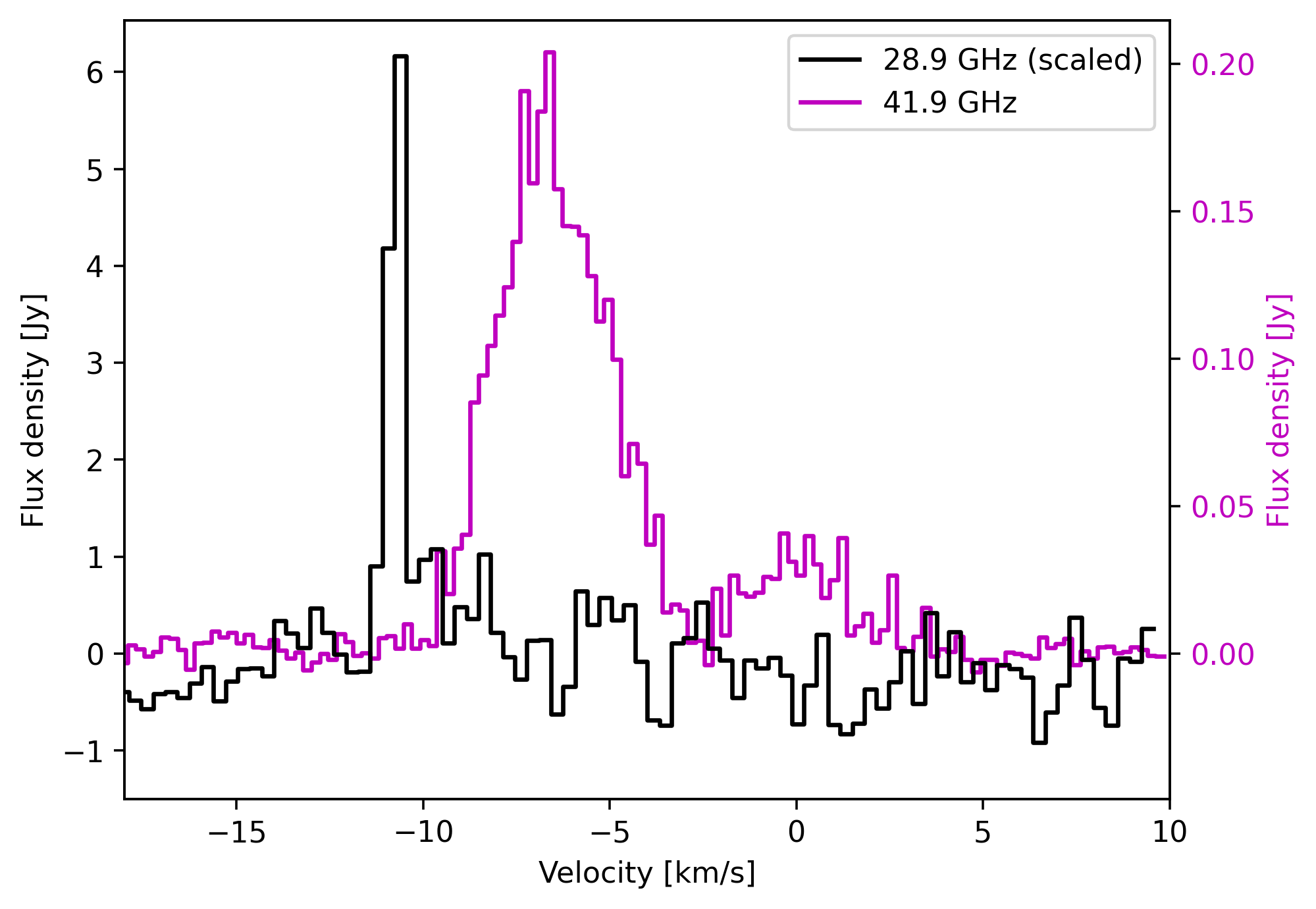}
\caption{NGC6334I Spectra Comparison; The signal in black indicates the 28.9 GHz line detected on 2019 April 30. Due to the difficulty in calibration of this transition, the amplitude of this signal has been scaled based on the RMS noise compared to the same transition observed at the same time toward G358.93‑0.03 (see Section~\ref{sec:ngc6334 results}). We estimate this gives an uncertainty of a factor of $\sim$2. The peak based on this calculation is $\sim$6 Jy. The signal in magenta is the 41.9 GHz line detected on 2019 April 26 with a peak of 0.2 Jy. The shape of this signal shows the characteristics of a thermal line and closely resembles the thermal detection toward this source in \cite{Wu:2023}. Note the difference in the scale of the respective axes; the 28.9 GHz line, as scaled by the procedure described in Section~\ref{sec:ngc6334 results}, is $\sim$30 times stronger than the 41.9 GHz line. The similarity in signal-to-noise despite the different scales is due to the longer integration time and the better bandpass in the 41.9 GHz observations}
\label{ngc6334_comparison_figure}
\end{figure}

Toward NGC6334I we detected the 28.9 GHz line at -10.5 km/s (Figure~\ref{ngc6334_comparison_figure}), which aligns with the velocity of the strongest class II methanol maser emission in NGC6334I-MM3 (at 6.7 GHz; \citealt{Ellingsen:2002}). Unfortunately due to the aforementioned difficultly of calibration at this frequency (see Section~\ref{sec:observations methods}), the low amplitude of the signal and the short integration time, bandpass and flux density calibration did not improve the quality of the data and often made it worse. We are unsure of why this level of difficulty in calibration did not affect the calibration of the 28.9 GHz line toward G358.93-0.03, as these observations were done during the same observing session. Therefore, for the 28.9 GHz transition in NGC6334I we instead scaled the amplitude of the signal by comparing the RMS noise of the uncalibrated data to the RMS noise of the equivalent result of the 28.9 GHz transition in G358.93‑0.03. As these two sources were observed during the same interleaved observations, we assume the system performance of the telescope is similar across these observations and therefore this calibration approach of scaling using noise can be used. While this is not an ideal calibration approach, none of the standard methods work for these data. 

We calculated the final scaling factor by taking the square root of the ratio of the integration times (as noise decreases with the square root of integration time) multiplied by the ratio of the RMS noise of a signal free area of each observation. Using the information in Table~\ref{full transition results} for the $S_{rms}$ and integration time of the observations toward G358.93‑0.03, along with the integration time of the observations toward NGC6334I and the original $S_{rms}$ of 0.0025 Jy, this scaling factor was determined to be $\sim$168 for the 28.9 GHz maser shown in Figure~\ref{ngc6334_comparison_figure}. We estimate this gives a large uncertainty of about a factor of 2, which means accurate amplitude measurements cannot be derived but we can infer the order of magnitude of the signal which helps in determining if this is a maser detection.

The 28.9 GHz line toward NGC6334I was detected in \cite{Wu:2023}, where the spectra shows a single spike along with a broadened, possibly thermal component. In \cite{Wu:2023} this signal was deemed a maser candidate. Our detection lacks the broadened element and has a much clearer spike. Our observations also show higher signal-to-noise, increasing the confidence that this is a maser detection. \cite{Wu:2023} observed the 28.9 GHz line on 2019 December 22, 8 months after our observations, which may explain the difference in properties of the observations as NGC6334I has been steadily decaying in maser intensity since the flare event was first reported in \cite{Hunter:2017}.

Our detection of the 41.9 GHz transition appears as a wider signal extending from -9.6 to +4.7 km/s with a peak flux density of 0.2 Jy. The thermal emission at 41.9 GHz in $^{13}$CH$_3$OH in \cite{Wu:2023} agrees nicely with our observations in terms of shape, width and approximate amplitude, the only major differences being the zoom band resolution and different date of our observations (0.224 km/s on 2019/04/26 for our observations and $\sim$0.65 km/s on 2019/06/11 for the observations in \citealt{Wu:2023}). The full properties of the detections toward NGC6334I are presented in Table~\ref{full transition results}.

\subsection{G345.01+1.79}
\label{sec:g345 results}

Toward G345.01+1.79 we report the first observations at 28.9 GHz and 41.9 GHz. A search across the entire band, with a focus on the region between approximately -25 km/s and -20 km/s, where most methanol masers seem to be concentrated in G345.01+1.79 \citep{Ellingsen:2012} showed no signal above 5$\sigma$ (where $\sigma$ = RMS noise level) in both 28.9 GHz and 41.9 GHz (5$\sigma$ = 9.8 Jy for 28.9 GHz and 0.04 Jy for 41.9 GHz). The full properties of the observations toward G345.01+1.79 are presented in Table~\ref{full transition results}.

\section{DISCUSSION}
\label{sec:discussion}

\subsection{G358.93-0.03}
\label{sec:g358 discussion}

Our observations toward G358.93-0.03 were taken during the same 2019 flare event observed in \cite{Miao:2022}, meaning that our observations were among the first detections of the 28.9 GHz line in this source. Additionally, the 41.9 GHz line of $^{13}$CH$_3$OH observed toward this source is the first ever detection of this transition as a maser. The combined observation of these two lines at 28.9 and 41.9 GHz in the same source also makes this the first isotopic maser detection of this methanol transition family.

\subsection{NGC6334I}
\label{sec:ngc6334 discussion}

Our observation of the 28.9 GHz line toward NGC6334I is the first confirmed detection of this line as a maser toward this source. Thermal emission from the 41.9 GHz line was reported previously by \cite{Wu:2023}, detected in observations 45 days after those that we present here. The shape and positions of the 28.9 GHz and 41.9 GHz lines in our data toward NGC6334I are noticeably different, with 28.9 GHz showing a thin maser line with a peak close to $\sim$11 km/s, while the 41.9 GHz signal is a broader thermal emission, centred closer to $\sim$7 km/s. This aligns well with previous class II maser and thermal detections toward NGC6334I (e.g., the comparison of multiple lines in \citealt{Cragg:2001}).

\subsection{G345.01+1.79}
\label{sec:g345 discussion}

Toward G345.01+1.79 we recorded the first observations at the frequencies of the 28.9 GHz and 41.9 GHz lines. As mentioned in Section~\ref{sec:g345 results}, previous detections of methanol masers toward G345.01+1.79 have been in the velocity range of -25 km/s to -20 km/s \citep{Ellingsen:2012}, but we found no detectable signal in this region or in the rest of the band. This result is still significant as it fills in a gap in observations at 28.9 GHz as seen in Table 1 of \cite{Ellingsen:2012}.

\subsection{Comparison of G358.93-0.03 Results to \cite{Miao:2022}}
\label{sec:miao comparison}

The observations in \cite{Miao:2022} toward G358.93-0.03 used the Tianma 65-m Radio Telescope (TMRT), first in the wideband mode on 2019 April 1 and then in later follow-up observations using the zoom-band mode on 2019 May 3.
The flux density of the 28.9 GHz line in \cite{Miao:2022} reaches a peak of 23.45 Jy at a velocity resolution of 0.01 km/s in the observations on 2019 May 3, this is an at-least twofold increase in flux density from our observation 4 days earlier on 2019 April 30. This difference may be explained in a range of different ways; differences between the telescopes and their observational parameters (e.g., effects of the antenna equipment on the signal, differences in sensitivity, etc.), the aforementioned problem of the 28.9 GHz line being out of the optimal frequency range of the ATCA 7mm receiver, and the rapid fluctuations of maser intensity over time during the flare event.

In \cite{Miao:2022} the 28.9 GHz line reached a peak of 14.24 Jy, with a wideband resolution of 0.95 km/s during the earlier observations on 2019 April 1. To compare the amplitude change over time \cite{Miao:2022} reduced the velocity resolution of the zoom-band detection on 2019 May 3 to match the wideband observation (from 0.01 km/s to 0.95 km/s). This shows that the peak amplitude changes from 14.24 Jy on 2019 April 1 to 10.68 Jy on the 2019 May 3. When the same resolution change is applied to our data, the peak amplitude is $\sim$7.6 Jy on 2019 April 30, showing some consistency with their results given the rapid changes with time and the uncertainties in amplitude calibration.

\subsection{Importance of Isotopic Maser Detections}
\label{sec:isotopic importance}

Detections of isotopic methanol masers, such as those detected here, can be used in modelling to more tightly constrain the physical properties of a star formation source. As shown in Figure 5 of \cite{Cragg:2001}, if multiple masers are observed toward the same source, the brightness temperature and gas density can be narrowed down based on `density dependence' graphs that have been produced from models based on existing maser detections. \cite{Cragg:2005} expands upon this by modelling brightness temperature with respect to gas temperature, dust temperature, gas density and specific column density. Therefore, knowing the lower limit on the brightness temperature of multiple transitions can greatly constrain these parameters in a maser model, making it possible to ascertain many source parameters from a few specific measurements. However, the fact that our observations were taken 4 days apart during a highly dynamic flare event (Section~\ref{sec:g358 results}) must be kept in mind when comparing the relative flux densities of our detections. Additionally, the comparison of relative flux densities may not be useful if the locations of the masers are shown to be non co-spatial (see Section~\ref{sec:spatial info}). Although there are a number of models of CH$_3$OH class II masers, there is limited modelling for these masers in $^{13}$CH$_3$OH. Therefore, while our observations provide new detections that can be used in maser modelling, it is beyond the scope of this paper to make assumptions about how these new detections may affect these constraints gained from this modelling.

\subsection{Rarity of the $8_2 \rightarrow 9_1  A^{-}$ Methanol Transition Family}
\label{sec:transition rarity}

Despite the rarity of the $8_2 \rightarrow 9_1  A^{-}$ methanol transition as a maser, results from the maser models in \cite{Cragg:2005} suggest that the 28.9 GHz transition theoretically has a high brightness temperature in a wide range of scenarios. This suggests that although the 28.9 GHz transition has rarely been observed as a maser, it may actually be more common than the lack of reported detections suggests. The frequency range from 26 to 30 GHz has historically not been available on many instruments. For example, in the ATCA, 28.9 GHz is precisely in the frequency gap between the K band (15mm, 16 to 25 GHz) and the Q band (7mm, 30 to 50 GHz) (ATCA Users Guide Table 1.1).

\subsection{Spatial Information for Each Transition}
\label{sec:spatial info}

Spatial information for these transitions was derived by imaging the data in MIRIAD. However, due to the aforementioned difficulties in reducing the data, the positional uncertainties are quite large. Therefore, the derived spatial information is discussed here with reference to this possibly large uncertainty.

For G358.93-0.03, imaging shows the 28.9 GHz maser to be within $\sim$0.6 arcseconds of the centre of the millimetre continuum source MM1 (centred at $\alpha$ = 17:43:10.1015, $\delta$ = -29:51:45.6936 (J2000); \citealt{Brogan:2019}), and shows the 41.9 GHz maser to be within 0.05 arcseconds of the centre. However both of these measurements are close or less than the scale of the positional uncertainty of the ATCA (0.4$''$; \citealt{Caswell:2009}). Therefore while these masers are co-spatial within the uncertainties of the data, the relative positions cannot be compared to the degree of high-resolution VLA observations \citep{Chen:2020a} or other VLBI observations \citep{Burns:2023}. The uncertainties in the rest frequencies for each transition result in line-of-sight velocity uncertainties of 0.5 km/s for the 28.9 GHz line and 0.2 km/s for the 41.9 GHz line. The position of the peaks for each line in our G358.93-0.03 data (Figure~\ref{g358_comparison_figure}) are separated by 1 km/s, suggesting that these two lines are coming from different regions with distinct line-of-sight velocities. In \cite{Chen:2020a} the $^{13}$CH$_3$OH transitions were detected in distinctly different areas from the CH$_3$OH transitions within the star forming region toward G358.93-0.03, both spatially and in line-of-sight velocity. Although we are not able to use positional data to determine if they are co-spatial, the differences in the velocities of the peak emission in each transition implies that the masers are also originating from distinct regions in our data.

For NGC6334I, imaging shows the 28.9 GHz maser appears to be close to NGC6334I-MM1 (centred at $\alpha$ = 17:20:53.415, $\delta$ = -35:46:57.88 (J2000); \citealt{Brogan:2016}), within $\sim$0.5 arcseconds of where 6.7 GHz masers have been detected \citep{Hunter:2021}. Previously observed maser emission in this area east of MM1 (increasing RA) usually have a line-of-sight velocity of $\sim$-7km/s or higher, whereas lower velocity maser emissions down to $\sim$-9.5km/s have been observed west of MM1 (decreasing RA) \citep{Hunter:2018, Kumar:2025}. Maser emission at the velocity of our 28.9 GHz result, at $\sim$-11km/s, have mostly been associated with the MM2 and UCHII MM3 regions of NGC6334I \citep{dePree:1995,Hunter:2018,Wu:2023,Kumar:2025}.

Due to this discrepancy in our data compared to previous observations, we used the continuum emission from our data (extracted via continuum subtraction in the same zoom band as the maser emission) to check the precision of our location of this line. As centimetre radio continuum emission is highly associated with the MM3 region in NGC6334I, if the imaging of the continuum correctly places it within MM3 we can have some confidence in our derived coordinates of the lines toward MM1. The imaging of the continuum did indeed place it in the MM3 region, approximately 4 arcseconds south of both MM1 \citep{Hunter:2021} and our line detections.

Imaging of the 41.9 GHz thermal line toward NGC6334I, also places it near MM1 in the northern area with components in the velocity range of -10 km/s to 4 km/s, which does align well with previous maser \citep{Hunter:2018} and thermal \citep{Wu:2023} detections in this area.

\subsection{Justifications for Maser Designations}
\label{sec:maser justification}

We can justify our designations of our maser detections by calculating their apparent brightness temperature. As astrophysical masers are characterised by having a strong signal at a specific frequency (due to population inversion into higher molecular states), calculating the brightness temperature of a sufficiently strong maser (here using the equation: $T = 1.77\times 10^{6}\frac{S_{\mathrm{Jy}}}{\nu^{2}_{\mathrm{Ghz}}d^{2}_{\mathrm{arcsec}}}$, where $S_{\mathrm{Jy}}$ is the flux density, $\nu$ is the rest frequency and $d$ is the angular size) will result in a brightness temperature above the usual maximum kinetic temperatures of thermal emission in the cores of massive young stellar objects (100 to 300 K; \citealt{Miao:2022}). Therefore, if the calculated brightness temperatures of our sources are higher than this threshold, it is likely that they are maser emission. 

In our observations, the synthesised beam sizes for both the 28.9 and 41.9 GHz observations are very large. (17.0$''$ x 2.3$''$ for 28.9 GHz and 7.7$''$ x 1.5$''$ for 41.9 GHz). Despite these maser sources likely being point sources within the ideal beam of the 7mm receiver of the ATCA (0.2$''$, ATCA Users Guide Table 1.1), we will conservatively use the minor axis of our large synthesised beam as the lower limit of the angular size. Toward G358.93-0.03 the calculated brightness temperatures are $\sim$4000 K for the 28.9 GHz line and $\sim$200 K for the 41.9 GHz line. From this alone we can see that the 28.9 GHz line is very likely a maser, however we will need further evidence of maser emission for the 41.9 GHz. 
The 41.9 GHz line toward G358.93-0.03 is even narrower than the 28.9 GHz, and both are especially narrow even in comparison to other maser lines in the source \citep{Breen:2019}, while the full width at half maximum (FWHM) of thermal lines are usually multiple km/s wide. From this evidence, we conclude that the 41.9 GHz line is also likely maser emission. Toward NGC6334I the calculated brightness temperatures are $\sim$2000 K for the 28.9 GHz line and $\sim$100 K for the 41.9 GHz line. This result along with the prominence of the signal for the 28.9 GHz line leads us to assess that this is indeed a maser, while the low brightness temperature and wide shape of the 41.9 GHz line leads us to conclude that this is most likely thermal emission.

\section{SUMMARY}
\label{sec:summary}

We report the first detection of the $8_2 \rightarrow 9_1  A^{-}$ 41.9 GHz maser in $^{13}$CH$_3$OH (methanol-13C) toward G358.93-0.03, along with the first confirmed maser detection of the same transition in CH$_3$OH (methanol) as a maser toward NGC6334I at 28.9 GHz. The first detections of these rare masers toward these sources presents more information for maser modelling, which can be used to constrain their known attributes.

The isotopic maser emissions in both our data and in \cite{Chen:2020a} appear to show that isotopic masers are strongest in different regions from the main species. Therefore future modelling of isotopic methanol masers may provide useful additional information on the physical conditions within high-mass star formation regions.

Additionally we add more temporal data for the 28.9 GHz line in G358.93-0.03 (first reported in \citealt{Miao:2022}) and the thermal 41.9 GHz line in NGC6334I, which can be used in analysis of the large changes over time during maser flaring events in each respective source. We also report the first observations (with no detections) for both transitions toward G345.01+1.79. This adds further information to this less studied southern-sky source.

\section*{ACKNOWLEDGEMENTS}
The Australia Telescope Compact Array is part of the Australia Telescope National Facility (grid.421683.a) which is funded by the Australian Government for operation as a National Facility managed by CSIRO. We acknowledge the Gomeroi people as the traditional owners of the Observatory site. This research has made use of NASA's Astrophysics Data System. We thank the anonymous referee for constructive feedback that has improved the paper.

\section*{DATA AVAILABILITY}
The data underlying this article will be shared on reasonable request
to the corresponding author. ATCA data are open access 18 months
after the date of observation and can be accessed using the Australia
Telescope Online Archive (\url{https://atoa.atnf.csiro.au}).




\bibliography{maser_paper_2025}


\end{document}